\documentclass[11pt,a4paper]{emulateapj}
\usepackage{natbib}
\usepackage{graphicx,color}
\usepackage{color}
\usepackage{amsmath}
\usepackage{epsfig}
\usepackage{epstopdf}
\usepackage{textcomp}
\usepackage{float}
\usepackage[english]{babel}
\usepackage[colorlinks=false]{hyperref} 
\usepackage{color}
\shorttitle{THE COLOR OF FRACTIONAL ANISOTROPY}
\shortauthors{FERNANDEZ, DOLE, ILIEV}
\begin{document}
\title{A Novel Approach to Constrain the Escape Fraction and Dust Content at High Redshift Using the Cosmic Infrared Background Fractional Anisotropy}%
\author{%
  Elizabeth R. Fernandez$^{1}$,  Herve Dole$^{1}$,  Ilian T. Iliev$^2$
}%
\affil{%
$^1$Institut d’Astrophysique Spatiale, Univ Paris-Sud 11 and CNRS, Orsay F-91405, France \\
$^2$Astronomy Centre, Department of Physics \& Astronomy, Pevensey II Building, University of Sussex, Falmer, Brighton BN1 9QH
}%
\email{
  $^1$Elizabeth.Fernandez@ias.u-psud.fr
}%

\begin{abstract}
The Cosmic Infrared Background (CIB) provides an opportunity to constrain many properties of the high redshift ($z>6$) stellar population as a whole.  This background, specifically from 1 to 200 microns, should contain information about the era of reionization and the stars that are responsible for  these ionizing photons.  In this paper, we look at the fractional anisotropy ($\delta I / I$) of this high redshift population, where $\delta I$ is the ratio of the magnitude of the fluctuations  and $I$ is the mean intensity.  We show that this can be used to constrain the escape fraction of the population as a whole, because the magnitude of the fluctuations of the CIB depends on the escape fraction, while the mean intensity does not.  This results in lower values of the escape fraction producing higher values of the fractional anisotropy.  This difference is predicted to be larger at longer wavelengths bands (above 10 microns), albeit it is also much harder to observe in that range. We show that the fractional anisotropy can also be used to separate a dusty from a dust-free population. Finally, we discuss the constraints provided by current observations on the CIB fractional anisotropy. 
\end{abstract}  
%%%%%%%%%%%%%%%%%
\section{INTRODUCTION}
\label{sec:introduction}
Modern cosmology is now able to constrain details about the era of reionization.  Observations show that the reionization of the universe occurred early and was extended in time, with an equivalent of an instantaneous reionization at $z \sim 11$ \citep{komatsu/etal:2008, komatsu/etal:2011}.  Stars are a likely candidate for being responsible for the majority of reionization because they are efficient producers of ultraviolet photons.  Thanks to a wave of modern, sensitive telescopes, we can begin to observe and understand the frontier of reionization, along with these high redshift stellar populations ($z>6$).  For example, we can observe these galaxies directly via high redshift surveys, which can now routinely identify a population of bright galaxies up to a redshift of about $z \sim 8$.  However, these surveys can only locate those galaxies that are both above the limiting magnitude and common enough to be present in the survey field.  It is now thought \citep{Bouwens/etal:2010,Robertson/etal:2010,fernandez/shull:2011} that reionization needed a large population of smaller galaxies below the current detection limits.  Because we cannot yet observe these galaxies directly, we can instead look for their cumulative light, which should exist as background radiation.  Because reionization is said to have occurred around $z \sim 11$, the photons responsible for reionization should be present in the Cosmic Infrared Background (CIB).  The spectral peak of this radiation is around the Lyman$-\alpha$ line, which will be redshifted to $1-4$ microns.  However, continuum emission will create an extended tail at longer wavelengths.  

Here, we discuss the CIB from about 1 to 200 microns.  The majority of the CIB will be emission from sources below $z \sim 6$, such as our Galaxy, foreground galaxies, and other sources of infrared light, such as zodiacal light.  If these sources can be subtracted away to a high precision, it is possible that the remainder could be from the era of reionization, and if so, could tell us about the properties of these high redshift stars.    

There have been many attempts to theoretically model the high redshift component of the CIB, especially in the near-infrared, from the mean \citep{santos/bromm/kamionkowski:2002, magliocchetti/salvaterra/ferrara:2003,salvaterra/ferrara:2003,
cooray/yoshida:2004,madau/silk:2005, fernandez/komatsu:2006}, to fluctuations \citep{kashlinsky/etal:2002,
kashlinsky/etal:2004,kashlinsky/etal:2005, 
kashlinsky/etal:2007, kashlinsky/etal:2012,
kashlinsky:2005,magliocchetti/salvaterra/ferrara:2003,
cooray/etal:2004,thompson/etal:2007a,
thompson/etal:2007b, fernandez/etal:2010, fernandez/etal:2011}.  
In this paper, we examine another way to analyze the CIB, the fractional anisotropy, which is the ratio of the fluctuations to the mean.  By looking at the fractional anisotropy, many free parameters are removed and more information about this elusive stellar population can be extracted.  Specifically, we discuss using the CIB as a probe for the escape fraction of ionizing photons.  Finding the escape fraction is important for understanding reionization and its duration.  There have been several attempts to measure the escape fraction through analytical models, simulations, and observations (see \citet{fernandez/shull:2011} and references within).   These papers have shown that the escape fraction appears to vary greatly from galaxy to galaxy.  Therefore, instead of trying to measure the escape fraction of an individual galaxy, here we discuss the average escape fraction of all galaxies, which will give more of a global view of reionization.  In addition, the fractional anisotropy can reveal information about the dust content of galaxies, which is mostly unknown at high redshifts.

We describe our simulations in section \ref{sec:sims} and our models in section \ref{sec:models}.  In section \ref{sec:anis}, we discuss our method for finding the mean CIB, the fluctuations of the CIB, and the fractional anisotropy.  In section \ref{sec:results}, we discuss our results of the fractional anisotropy for various bands.  In section \ref{sec:obs}, we discuss the most recent observations.  We conclude in section \ref{sec:conc}.  Throughout this paper, we use the cosmological parameters
($\Omega_{\rm m}$, $\Omega_\Lambda$, $\Omega_{\rm b}$, $h$)=(0.27, 0.73, 0.044, 
0.7), consistent with the simulations from  \citet{iliev/etal:2011}, which are based on the WMAP 5-year results and other
available constraints \citep{komatsu/etal:2008}.

\section{THE SIMULATIONS}
\label{sec:sims}

In order to predict the angular power spectrum of the CIB, we used simulations from \citet{iliev/etal:2011}, which are N-body simulations combined with radiative transfer, which allow us to see how sources are affected by the reionization process.  The high resolution of these simulations (with a minimum mass of $10^8 M_\sun$) allow us to also include Jeans-mass filtering on low mass halos.  This effectively allows suppression of star formation within small halos ($10^8-10^9 M_\sun$) because of elevated gas temperatures that could be caused by the proximity to other star forming galaxies.  These simulations have a box size of either $114  h^{-1}  Mpc$ (for cases with suppression of small sources) or $37 h^{-1}  Mpc$ (with no suppression, where $M_{min}=10^8 M_\odot$, or complete suppression, where $M_{min}=10^9 M_\odot$).  These simulations are summarized in Table \ref{tab:sims} \citep{iliev/etal:2011, fernandez/etal:2011}.  

To describe the stellar populations within these halos and their relationship with their environment, we define a parameter $f_\gamma$, which describes the number of ionizing photons produced per stellar atom that can escape the galaxy and reionize the intergalactic medium (IGM).  $f_\gamma$ is defined as a product of the star formation efficiency, or the fraction of baryons that are in stars ($f_*$), the escape fraction ($f_{esc}$), and the number of ionizing photons per stellar atom ($N_i$):  
\begin{equation}
\label{eq:fgamma}
 f_\gamma= f_{esc}f_*N_i.
 \end{equation}
We allow $f_\gamma$ to have different values, dependent on the mass of the halo, assuring it is consistent with reionization. 
 
%%%%%%%%%%
\begin{table*}[t]
\begin{center}
\begin{tabular}{|l|l|l|l|l|l|l|l|l|l|}

\hline
Simulation Name & Box Size &  Minimum  & Suppression & $f_{\gamma,large}$ 
& $ f_{\gamma,small}$ & $z_{\rm ov}$ & $\tau$  \\
& (Mpc) &  Halo Mass ($M_\sun$) & & & & &  \\
\hline
Partial Suppression, & & & & & & &\\
{High Efficiency} & 163  & $10^8$ & Yes  &  10  & 150 &8.3 &0.080\\
Partial Suppression, & & & & & & & \\
{Low Efficiency}   & 163  & $10^8$ & Yes  & 2  & 10 & 6.7&0.058\\
{No Suppression}   & 53 & $10^8$ & No  & 0.4  & 6 & 8.6 &0.078\\
{Complete Suppression} & 53  & $10^9$ & Yes - complete & 12 & 0 & 8.3 &0.071 \\
\hline
\end{tabular}

\caption{Radiative transfer simulations used in this work.  $z_{ov}$ is the redshift of overlap, where reionization is complete, and $\tau$ is the electron scattering optical depth.  $f_{\gamma,small}$ is for halos that are between $10^8$ and $10^9 M_\sun$ while $f_{\gamma,large}$ is for halos above $10^9 M_\sun$.
}%
\label{tab:sims}
\end{center}
\end{table*}
%%%%%%%%%%%%%%%%%%%%%%%%%%%%%%%%%%%%%%%%%%%%%%%%%%%%%%%%%%%%%%%%%%%%%%

\section{OUR MODELS}
\label{sec:models}
%There is a good deal of uncertainty to both the mass and metallicity of the stars at high redshift.  Therefore, we use a limiting cases for our models for both the metallicity and mass.  
True first generation stars are metal free (Population III stars).  As time goes on, stars die and enrich the universe, and eventually, these Population III stars give way to stars with metals (Population II stars).  It is unclear when this happens, and this process is probably very inhomogeneous.  Therefore, we assume two limiting cases - all of the stars from $6<z<30$ are either Population III ($Z=0$) or Population II ($Z=1/50  Z_\sun$) stars.

In addition to the metallicity, there is uncertainty for the mass spectrum of these stars.  These stars could be very large, or they could be similar in size to what we see today.  To model these two extremes, we choose either a heavy, Larson mass function \citep{larson:1998}:
\begin{equation}
f(m)\propto m^{-1}\left(1+\frac{m}{m_c}\right)^{-1.35},
\end{equation}
with mass limits of $m_1 = 3M_\sun$, $m_2=500M_\sun$, and $m_c = 250M_\sun$ to model a population of large stars, or a Salpeter mass function \citep{salpeter:1955}:
\begin{equation}
f(m) \propto m^{-2.35},
\label{eq:salpeter}
\end{equation}
with mass limits of $m_1=3M_\sun$ and $m_2=150 M_\sun$ to simulate a mass spectrum similar to what we see in the local universe.

If we combine our limiting cases for both mass and metallicity, we can establish our two limiting stellar models: Population III stars with a Larson mass spectra, and Population II stars with a Salpeter mass spectra.  In reality, these stellar limits are extreme.  In addition, we expect stellar properties to be inhomogeneous throughout redshift.   However, these examples were chosen as limiting cases: which represent a population with the smallest and largest amplitude for the angular power spectrum of a large range of models, studied in detail in \citet{fernandez/etal:2010}.  We would expect the actual amplitude for the angular power spectrum to lie between these extremes.  These populations are summarized in Table \ref{tab:pop}.

%%%%%%%%%%
\begin{table*}
\begin{center}
\begin{tabular}{|l|l|l|l|l|l|l|l|l|l|l|l|}
\hline
$f_{\gamma}$ & $ f_{\rm esc}$ & $f_{*}$ -  Pop II Salpeter & $f_{*}$ -  Pop III Larson \\
\hline
$10$   & $0.1$ &$ 3.8 \times 10^{-2}$ & $4.0 \times 10^{-3}$ \\
$10$  & $0.3$ &$ 1.3 \times 10^{-2}$ & $1.3 \times 10^{-3}$\\
$10$  & $0.5$ &$ 7.7 \times 10^{-3}$ & $8.0 \times 10^{-4}$ \\
$10$   & $1$ &$ 3.8 \times 10^{-3}$ & $4.0 \times 10^{-4}$\\
  $ 150$   & $0.1$ & $5.8 \times 10^{-1}$ & $6.0 \times 10^{-2}$ \\
  $150$   & $0.3$ &$ 1.9 \times 10^{-1}$ & $2.0 \times 10^{-2}$\\
$150$   & $0.5$ &$ 1.2 \times 10^{-1}$ & $1.2 \times 10^{-2}$\\
$150$   & $1$ &$ 5.8 \times 10^{-2}$ & $6.0 \times 10^{-3}$\\
 \hline
\end{tabular}

\caption{The properties of the stellar populations.  $f_*$ was set to be consistent with reionization.}%
\label{tab:pop}
\end{center}
\end{table*}
%%%%%%%%%%%%%%%%%%%%%%%%%%%%%%%%%%%%%%%%%%%%%%%%%%%%%%%%%%%%%%%%%%%%%%

\section{COMPUTING THE FRACTIONAL ANISOTROPY}
\label{sec:anis}
\subsection{The Mean Cosmic Infrared Background}
\label{sec:mean}

Now that we have our stellar and galactic models, we are in a position to calculate the fractional anisotropy, $\delta I / I$.  To do this, we must compute both the mean intensity and the angular power spectrum of the CIB.  The mean CIB is a combination of emission from the star, which is modeled as a stellar blackbody, and emission from the nebula, which is a combination of the Lyman$-\alpha$ line,   two-photon, free-free and free-bound emission, and with the possibility of emission from dust (see section \ref{sec:dust}).  This nebular emission is either produced within the halo itself, or within the IGM if some fraction of the ionizing radiation ($f_{esc}$) escapes the halo.  However, the mean CIB does not depend on $f_{esc}$, since the nebular emission is the same, regardless of whether it is from the halo or the IGM.

Each emission process (stellar, free-free, free-bound, two-photon, and the Lyman$-\alpha$) was modeled analytically (for details, see \citet{fernandez/komatsu:2006}) and integrated over a range of redshifts from $6<z<30$.  The total intensity (I) is then:
\begin{eqnarray}
\label{eq:mean}
\nonumber
 I &=& \frac{c}{4\pi}\left(f_*\frac{\Omega_b}{\Omega_m}\right)\int
  \frac{dz}{H(z)(1+z)}  \bar{\rho}_M^{halo}(z)\\
& &\times
\left[\bar{l}^*(z) +\bar{l}^{ff}(z)+\bar{l}^{fb}(z)+\bar{l}^{2\gamma}(z)+\bar{l}^{\rm Ly\alpha}(z)\right].
\end{eqnarray}
Here, $\bar{\rho}_M^{\rm halo}(z)$ is the mean mass density collapsed into halos from the simulation.  The luminosity per stellar mass, $\bar{l}(z)$, is given for each component of the luminosity, $*$ for stellar, $ff$ for free-free, $fb$ for free-bound, $2\gamma$ for two-photon, and $Ly\alpha$ for Lyman-alpha emission.  The luminosity of any component "$\alpha$" can be written as\footnote{This expression is only valid if the average stellar lifetime is always less than the star formation time scale $t_{SF}$, which is true for the cases we are concerned with.  For more information, see \citet{fernandez/etal:2010}.} :
\begin{equation}
 l^\alpha_\nu(z) = \frac{d\ln\rho_*(z)}{dt} 
\frac{\int^{m_2}_{m_1} dm
   f(m)L^\alpha_{\nu}(m)\tau(m)}{\int^{m_2}_{m_1} dm f(m) m}.
\end{equation}
The luminosity of each emission component ($L_\nu^\alpha$),  and the stellar lifetime ($\tau(m)$) are integrated over a mass spectrum of stars.  The first part of this expression is the inverse of the star formation time scale, $t_{SF}(z)=[\frac{d\ln\rho_*(z)}{dt}]^{-1}$.  This star formation time scale is unknown, but we assume a value of $11.5 \; \rm{Myr}$, consistent with the value from simulations of \citet{iliev/etal:2011}.  %Because the average stellar lifetime is always less than $t_{SF}$, 
This expression then reduces to:
\begin{equation}
 l^\alpha_\nu(z) = \frac{1}{t_{SF}(z)} 
\frac{\int^{m_2}_{m_1} dm
   f(m)L^\alpha_{\nu}(m)\tau(m)}{\int^{m_2}_{m_1} dm f(m) m}
\end{equation}
\citep{fernandez/etal:2010}.{\footnote{The value chosen for $t_{SF}$ will change the amplitude of the luminosity, which will both affect the mean and the fluctuations of the CIB.  A benefit of taking the fractional anisotropy is that the dependence on $t_{SF}$ will nearly cancel out.  For more information on the dependence of the luminosity on $t_{SF}$, see section 6.1 of \citet{fernandez/etal:2010}.}}.   

\subsection{Dust}
\label{sec:dust}
We do not know how much dust exists in high redshift galaxies.  Molecular gas is already observed at $z\sim5$, an indication that dust is present at those redshifts \citep{riechers/etal:2010}.  Because dust will affect the spectra of high redshift galaxies, the fractional anisotropy may also change.  In order to see if our results are affected by dust, we compute the spectra expected if the radiation field is further reprocessed by a dusty medium.  We generated a dust spectrum using DustEM \citep{compiegne/etal:2011} predicted for a galaxy with a high metallicity and minimal destruction of dust grains \citep{compiegne/etal:2010}. In reality, the low metallicity and hard radiation fields expected at high redshift will lead to a dust contribution that is less than the one modeled here. In addition, DustEM models are computed in the optically thin limit, so therefore, the dusty SED we obtain is the upper limit for the amount that dust will redden. Our dusty model represents the extreme model for a dusty galaxy, with our case with no dust representing the opposing limit. 

%
%We generated a dust spectrum using DustEM } predicted for a galaxy with a high metallicity and minimal destruction of dust grains .  The model presented is the optically thin limit.  In reality, the low metallicity and hard radiation fields expected at high redshift, as well as having an optically thick medium, will lead to a dust contribution that is less than the one modeled here.  Therefore, our model with dust will represent the extreme limit of a dusty galaxy.  Our model with no dust represents the opposite extreme.  

%In addition, DustEM models the optically thin limit, so therefore, the dusty SED we obtain is the upper limit for the amount that dust will transform the SED for the wavelength range we are concerned with.  Our dusty model represents the extreme model for a dusty galaxy, with our case with no dust representing the opposing limit. 

\subsection{Fluctuations in the Cosmic Infrared Background}
\label{sec:fluc}

The next step is to compute the angular power spectrum.  These fluctuations will arise from both the emitting halos and their surrounding HII regions within the IGM.  As shown in \citet{fernandez/etal:2010}, the fluctuations from the IGM are probably quite small (from 2 to 7 orders of magnitude smaller than that of the halos themselves) so can safely be ignored.  

The angular power spectrum $C_l$ can then be written as:  
\begin{eqnarray}
\label{eq:CL}
\nonumber
  C_l &=& \frac{c}{(4\pi)^2}\left(f_*\frac{\Omega_b}{\Omega_m}\right)^2\int
  \frac{dz}{H(z)r^2(z)(1+z)^4}\\
\nonumber
& &\times\left[\bar{\rho}_M^{halo}(z)
\left\{\bar{l}^*(z) +
 (1-f_{\rm
 esc})\bar{L}(z)\right\}\right]^2\\
& &\times
b^2_{eff}\left(k=\frac{l}{r(z)},z\right)P_{\rm
  lin}\left(k=\frac{l}{r(z)},z\right).
\end{eqnarray}
Here $b_{eff}$ is the effective bias,  $P_{\rm lin}$ is the linear matter power spectra, $r(z)=c\int^z_0 dz'/H(z')$ is the comoving distance, and the luminosity is:
\begin{equation}
\bar{L}(z)=\bar{l}^{ff}(z)+\bar{l}^{fb}(z)+\bar{l}^{2\gamma}(z)+\bar{l}^{\rm
       Ly\alpha}(z).
\end{equation}
The simulations provide both the halo bias and the linear
matter density fluctuations.  Note that the angular power spectrum depends on $f_{esc}$.  The angular power spectrum of these simulations was computed in Fernandez et al. 2012.

\subsection{The Fractional Anisotropy}
\label{sec:fracanis}

The fractional anisotropy  of the CIB is obtained by dividing the angular power spectrum $C_l$ (shown in equation \ref{eq:CL}) by the mean intensity $I$ (equation \ref{eq:mean}):
\begin{equation}
\delta I/I\equiv \sqrt{l(l+1)C_l/(2 \pi I^2)}.
\label{eq:fracanisdef}
\end{equation}
Most of the free parameters then cancel out, including the star formation efficiency $f_*$.  The luminosity $\bar{l}^\alpha$, however, will only cancel out when $f_{esc}=0$. %  almost cancels out (this dependence will be discussed more in the next section).  However, $f_{esc}$ does not.  
Therefore, the fractional anisotropy serves as a test to constrain $f_{esc}$.

\section{RESULTS}
\label{sec:results}
%\subsection{Fractional Anisotropies for Various Escape Fractions}
At large values of $l$, the minimum mass of the star forming halos and suppression history will change the shape of the angular power spectrum due to non-linear bias effects \citep{fernandez/etal:2011}.  Since the minimum mass of these star forming halos is unknown, we compute the fractional anisotropy for $l=3000$, avoiding any flattening or steepening of the angular power spectrum that could occur at larger l.  

\begin{figure*}
\centering \noindent
\includegraphics[width=12cm]{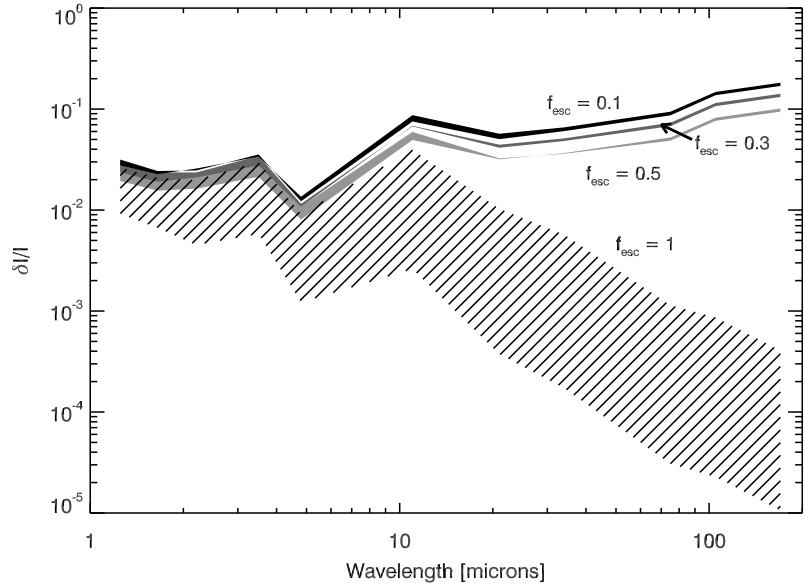}
\caption{The Fractional Anisotropy, $\delta I / I$, as a function of wavelength, for various values of $f_{esc}$, in cases without dust.  Population II stars with a Salpeter mass spectrum provide the upper limit of the shaded regions, while Population III stars with a Larson mass spectrum provide the lower limits.  Other reasonable assumptions for the mass and metallicity of the stellar populations should lie within the shaded regions.   }
\label{fig:Frac1}
\end{figure*}

\begin{figure*}
\centering \noindent
\includegraphics[width=12cm]{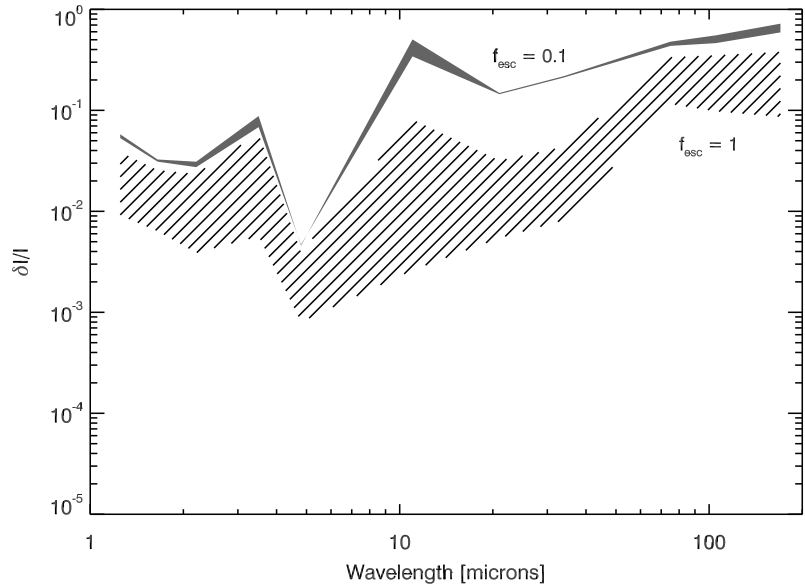}
\caption{The Fractional Anisotropy, $\delta I / I$, as a function of wavelength, for cases with dust. }
\label{fig:Frac3}
\end{figure*}

The fractional anisotropy at $l=3000$ is shown in Figures \ref{fig:Frac1} and \ref{fig:Frac3} as a function of observed wavelength.  The shaded regions are bounded by our two fiducial models (Population II stars with a Salpeter mass function will give the upper limit of the shaded region, while Population III stars with a Larson mass function will give the lower limit).  Other reasonable models, varying the mass or metallicity of the stars, will lie between these two limiting cases, since the amplitude of the angular power spectrum will lie between these cases \citep{fernandez/etal:2010}.  We also show a range of $f_{esc}$, from $f_{esc} = 1$, (where all the ionizing photons escape from the halo into the IGM), to $f_{esc}=0.1$.  Results are shown for a case where reionization progresses with a high efficiency, the minimum halo mass is $10^8 M_\odot$, and small halos can be suppressed.  However, these assumptions do not greatly affect the results.  

As seen in this figure, the escape fraction has a large effect on $\delta I / I$.  The mean level of the CIB, given in equation \ref{eq:mean}, has no dependence on the escape fraction.  The angular power spectrum, given by equation \ref{eq:CL}, has a factor of $(1-f_{esc})$.  Therefore, when $f_{esc}$ rises, the level of the nebular contribution to the angular power spectrum will fall.  This causes the overall level of %
%
%In both figure \ref{fig:Frac1} and \ref{fig:Frac3}, as $f_{esc}$ increases, 
$\delta I / I$ to fall.  

\begin{figure*}
\centering \noindent
\includegraphics[width=12.3cm]{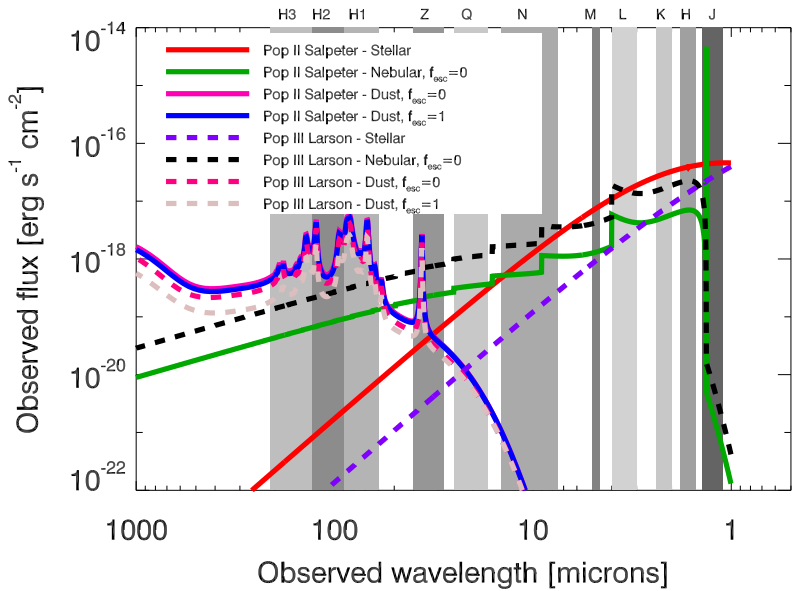}
\caption{The spectra of a $10^9  M_\odot$ galaxy at $z=10$.  We assume a star formation efficiency of $f_*= 0.1$.  Two stellar populations are shown - Population II stars with a Salpeter mass spectrum, or Population III stars with a Larson mass spectrum.   In addition, two limiting cases of the escape fraction are shown - $f_{esc} = 0$ and $f_{esc}=1$.  When $f_{esc} =0$, the ionizing radiation stays within the galaxy, creating nebular emission, which is not present in the halo if $f_{esc} =1$.  The dust emission will fall only slightly when $f_{esc}=1$.    Bands are denoted by the shaded gray regions.  Hershel PACS bands are labeled as H1, H2, and H3.  }
\label{fig:spectra}
\end{figure*}

%
%This difference is the most pronounced for wavelengths greater than about 10 microns.  
This drop-off of $\delta I / I$ for larger values of the escape fraction is more pronounced at longer wavelengths.  
To see why this occurs, we can look at the mean spectrum of starlight and nebular emission for a high redshift galaxy in Figure \ref{fig:spectra}.  (The definition of the bands shown are given in Table \ref{table:bands}.)  In the near-infrared bands ($\lambda <4\; \mu m$), there is always a large contribution from the stellar blackbody emission.  At longer wavelengths, the stellar emission drops off very quickly, while, if the escape fraction is low, the nebular emission remains relatively high.  If the escape fraction is high, however, the nebular emission component of the angular power spectrum would be diminished.  This is  particularly noticeable at longer wavelengths.    

%
%
%For metal poor Population II stars, the stellar blackbody is always greater than the nebular component for the near-infrared wavelengths .  As the wavelength increases, the emission from the stellar component will drop below the level of emission from the nebula.  This happens in the N band for Population II stars.  For Population III stars, the blackbody spectrum of the star is steeper, since these massive, metal free stars have a harder spectra.  As a consequence of the stellar blackbody being shifted to higher energies and because there are more ionizing photons that will be processed into nebular emission, the stellar spectrum is almost always equal to or less than that of the nebular component.  In the far-infrared bands, the stellar component is quite low.
%
%This behavior of the emission, coupled with the escape fraction, will affect the measured value of $\delta I / I$.  When the escape fraction is very high, the nebular contribution of the halo angular power spectrum will fall to zero, while the mean will remain the same.  Because of this, $\delta I / I$ will have smaller values for larger values of $f_{esc}$.  For both metallicities of stars, there is still a significant amount of stellar emission in the near-infrared bands.  % so both the mean and the angular power spectrum will be higher, therefore, $\delta I / I$ will be high as well.  
%However, for longer wavelengths, the stellar contribution becomes negligible.  If $f_{esc}$ is high, the amplitude of the angular power spectrum will drop in relation to the mean, so $\delta I / I$ drops to be low.  

For small values of $f_{esc}$, the range of allowed values for $\delta I / I$ is quite narrow.  This range widens if $f_{esc}=1$.   This is a consequence of the spectral shape for our two stellar models.  For Population III stars, the blackbody spectrum of the star is steeper and there are more ionizing photons to be processed into nebular emission.  The stellar spectrum is almost always equal to or less than that of the nebular component.  On the other hand, the stellar blackbody is greater than the nebular component at short wavelengths for metal poor Population II stars.  As the wavelength increases, the emission from the stellar component will drop below the level of emission from the nebula.  If $f_{esc}$ is small, the total emission of the halos for Population III and Population II stars is similar, so the range of $\delta I / I$ is narrow.  If $f_{esc}$ is large, the amplitude of the angular power spectrum of large Population III stars will be more affected than smaller Population II stars, widening the range of allowed values of $\delta I / I$.

%This is because when $f_{esc}=1$, the difference between our fiducial models is the greatest.  When $f_{esc}=0$, the luminosity of each emission component cancels out completely in equation \ref{eq:fracanisdef}.  As $f_{esc}$ increases, the luminosity of the nebular component no longer completely cancels out.  Rather, the luminosity of the nebular component of the angular power spectrum will fall, diminished by a factor of ($1 - f_{esc}$), while the nebular component of the mean will remain the same. Therefore, the difference between each model with a different mass and metallicity, which in turn affects the luminosity, will increase, widening the range of allowed values for $\delta I/I$.

%
%%%%%%%%%%%%%%%%%%%%%%%%%%%%%%%%%%%%%
%%%%%%%%%%%%%%%%%%%%%%%%%%%%%%%%%%%%%
%%%%%%%%%%%%%%%%%%%%%%%%%%%%%%%%%%%%%
%%%%%%%%%%%%%%%%%%%%%%%%%%%%%%%%%%%%%
%%%%%%%%%%%%%%%%%%%%%%%%%%%%%%%%%%%%%
%%%%%%%%%%%%%%%%%%%%%%%%%%%%%%%%%%%%%

%%%%%%%%%%%%%%%%%%%%%%%%%%%%%%%%%%%%%
\begin{table}[t]
\begin{center}
\begin{tabular}{|l|l|l|l|l|l|l|l|l|l|}
\hline
Band & Central of Band ($\mu m$) & Waveband ($\mu m$)  & $\Delta \lambda / \lambda$ \\
\hline
J & 1.25  & 1.1 - 1.4 & 24\% \\
H & 1.65  &  1.5 - 1.8 & 18\% \\
K & 2.2  & 2.0 - 2.4 & 18\% \\
L & 3.5  & 3.0 - 4.0 &29\% \\
M & 4.8  & 4.6 - 5.0 &8.3\%\\
N & 11 & 7.5 - 14.5 & 64\%\\
Q & 21  & 17  - 25 &38\%\\
Z &  34  &  28 - 40 &35\%\\
H1 & 75  &  60 - 90 &40\%\\
H2 &110  &   90 - 130 &36\%\\
H3  & 170  &  130 - 210 & 47\%\\
\hline
\end{tabular}
\caption{%
   Band definitions used for infrared bands.  H1, H2, and H3 denote the bands of the Herschel PACS instrument.  All bands are assumed to be rectangular. 
}%
\label{table:bands}
\end{center}
\end{table}
%%%%%%%%%%%%%%%%%%%%%%%%%%%%%%%%%%%%%

In Figure \ref{fig:Frac3}, we see the the fractional anisotropy for the case when galaxies contain dust.  Here, we no longer see a decrease in $\delta I / I$ at long wavelengths.  This is because dust will reprocess the stellar and nebular emission, re-emitting the light at longer wavelengths.  (An illustration of the SED with dust included is also shown in Figure \ref{fig:spectra}.)  Unlike nebular emission, the dust component will not fall to zero when $f_{esc}=1$.  In fact, the change between the dust emission from $f_{esc}=0$ and $f_{esc}=1$ is only slight.  Therefore, while the nebular component will disappear if $f_{esc}=1$, the dust component will still be present,  causing the angular power spectrum at long wavelengths to remain high, raising $\delta I / I$.  This is a direct result from the fact that nebular emission only results from ionizing photons, while photons of lower energies can be converted into dust emission.

\section{COMPARISON TO OBSERVATIONS}
\label{sec:obs}

Measurements of the CIB are notoriously hard to perform.  Finding an accurate mean is particularly difficult because precise foreground subtraction is needed.  However, observations, such as with CIBER, AKARI, and Herschel, continue to improve, resolving foregrounds in more detail and obtaining more reliable observations for the CIB.

Many observations have been made to try to understand the contribution of high redshifts to the CIB from 1-4 microns.  In order to uncover any residual emission in the mean or fluctuation observations, one must carefully take into account all of the foreground components.  Zodiacal light is a major contaminant, and because it is very difficult to model, it is not straightforward to subtract from the CIB.  In addition, foreground galaxies at lower redshifts must be taken into account.  
Despite the difficulty, there have been many attempts to measure the mean level of the CIB in the near-infrared due to high redshift stars \citep{dwek/arendt:1998,gorjian/wright/chary:2000,kashlinsky/odenwald:2000,wright/reese:2000,
wright:2001,cambresy/etal:2001,totani/etal:2001, kashlinsky/etal:2002,kashlinsky/etal:2004, kashlinsky:2005, kash/etal:2007, kashlinskyb/etal:2007c, kashlinsky/etal:2012, magliocchetti/salvaterra/ferrara:2003, odenwald/etal:2003,cooray/etal:2004, matsumoto/etal:2005, thompson/etal:2007a,thompson/etal:2007b}.   

Fluctuation observations are, in theory, easier to perform, since they do not need an accurate zero point, and instead rely on variations from one region of the sky to another.  However, these observations still rely on careful and complete subtraction of foreground sources, and also remain controversial \citep{kashlinsky/etal:2005, kashlinskyb/etal:2007c, kashlinsky/etal:2012, cooray/etal:2007, thompson/etal:2007a, matsumoto/etal:2011}.
   
Observations are even more difficult in the mid and far-infrared.  One problem is that foregrounds that were present in the near-infrared are even more prevalent in the mid and far-infrared.  Zodiacal light peaks at about 20 $\mu m$, which washes out most detections of the CIB in this range.  In addition, Galactic cirrus is a main contaminant, however, observations are possible in clean regions of the sky.  Finally, as wavelength increases, foreground galaxies become more difficult to resolve.  All of these problems lead to only a fraction of the CIB in the mid and far-infrared being resolved into  low-redshift galaxies.  It is likely that only a very small (and currently unknown) percentage of this excess is from $z>6$, 
so care must be taken in interpreting observations.
  
There has been a great push to understand the CIB at longer wavelengths.  At 100 and 160 $\mu m$, \citet{Penin/etal:2011}  measured the mean and fluctuation power of galaxies at all redshifts using observations from IRIS/IRAS and Spitzer/MIPS, respectively.  Fluctuations of the cumulative CIB have been taken in the mid-infrared to submillimeter wavelengths \citep{kashlinsky/odenwald:2000,Lagache/Puget:2000, Miville-Deschenes+2002,Grossan/smoot:2007, Lagache/etal:2007,Amblard/etal:2011,matsuura/etal:2011,Planck:2011, pyo/etal:2012}.  
The cumulative mean level of the CIB from galaxies at all redshifts 
has been measured as well \citep{fixsen/etal:1998,  Hauser/etal:1998,  Lagache/etal:2000, wright:2004,odegard/etal:2007, matsuura/etal:2011}.
The mean CIB has been measured as a function of redshift 
\citep{Berta/etal:2011,Jauzac/etal:2011,Bethermin/etal:2012}, while measurements from BLAST and {\it{Planck}} from 250 to 1400 microns \citep{Planck:2011} 
could indicate that galaxies at a higher redshift (here, $z>1.2-2$) could contribute more to the CIB as the wavelength increases.  Currently, the best measurements \citep{Planck:2011, Viero/etal:2009} show that the fractional anisotropy is at the order of 15\%, however, these measurements include galaxies at all redshifts.

One way to subtract unresolved low redshift galaxies in the mid to far-infrared to a more complete level is to use a stacking algorithm.  This typically involves using the locations of known galaxies at a shorter wavelength, stacking these locations of a longer wavelength image, and utilizing this stack to calculate the CIB accounted from these  galaxies.   If stacking is relied upon, more of the CIB at long wavelengths can be resolved into lower redshift galaxies.  For example, \citet{Marsden/etal:2009} used stacking to resolve 100\% of the CIB as detected with FIRAS \citep{fixsen/etal:1998} at 250, 350, and 500 microns using BLAST.  \citet{Dole/etal:2006} used 24 $\mu m$ sources from Spitzer/MIPS data to stack images at 70 and 160 $\mu m$.  They were able to resolve 79\%, 92 \%, and 69\% of the CIB at 24, 70, and 160 $\mu m$ respectively.  \citet{Berta/etal:2010} resolved 45\% and 52\% (without stacking) and 50\% and 75\% (with stacking) of the CIB at 100 and 160 $\mu m$ using Herschel/PACS data.  At longer wavelengths, \citet{Greve/etal:2010} resolved 16.5\% of the CIB at 870 microns using stacking.  While it is possible that some of the remaining flux is from low redshift galaxies, it is also possible that some of this unresolved CIB could be due to high redshift galaxies.  (See, for example, \citet{matsuura/etal:2011}).  

\begin{figure*}
\centering \noindent
\includegraphics[width=12cm]{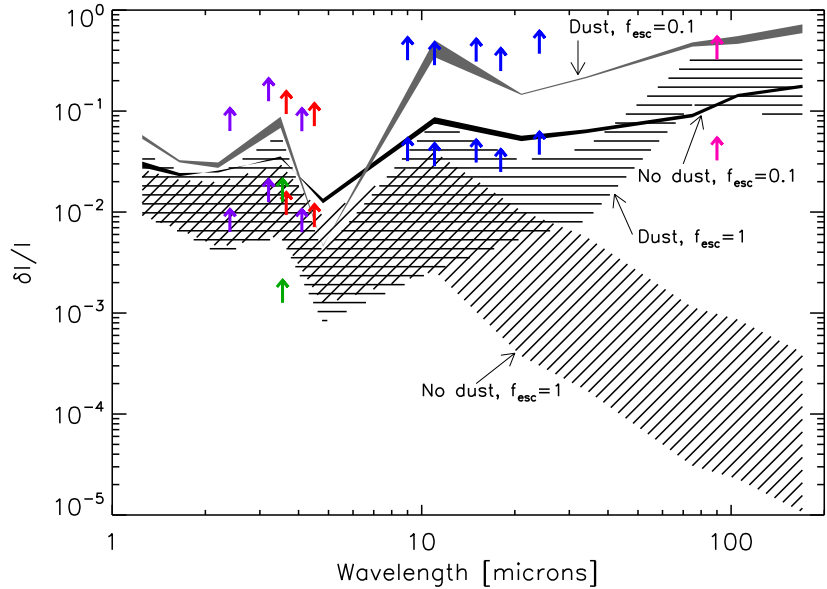}
\caption{The Fractional Anisotropy, $\delta I / I$, in comparison to recent observations.  The upper limit of the mean CIB from $z>6$ stars is assumed to be either $10$ nW m$^{-2}$ sr$^{-1}$ (lower set of arrows) or $1$ nW m$^{-2}$ sr$^{-1}$ (upper set of arrows).  These assumptions provide the lower limits of $\delta I / I$.  Shown are observations from Kashlinsky et al (2012) (red arrows), Matsumoto et al (2011) (purple arrows), Cooray et al (2007) (green arrows), Pyo et al (2012) (blue arrows), and Matsuura et al (2011) (pink arrows).  The horizontal hatched region shows cases with dust and $f_{esc}=1$, the diagonal hatched region shows cases with no dust and $f_{esc} = 1$, the grey shaded region shows cases with dust and $f_{esc} = 0.1$, and the black region shows cases with no dust and $f_{esc} = 0.1$.}
\label{fig:obs}
\end{figure*}

We compare some of these observations to our models for the fractional anisotropy at high redshifts.  While precise measurements of the mean are challenging, measurements of the fluctuation power are becoming more reliable.  Because of this, in Figure \ref{fig:obs} we show the fractional anisotropy predicted using various recent observations of the fluctuation power, assuming an upper limit of the mean CIB due to high redshift stars is either $10$ nW m$^{-2}$ sr$^{-1}$ or $1$ nW m$^{-2}$ sr$^{-1}$.  Because it is unlikely that a $z >6$ component of the mean CIB will be much higher than this, very low values of $\delta I / I$, and thus very high values of the escape fraction from a dust-free population, can be ruled out.  More definitive conclusions can be reached as observations continue to improve and our understanding of foreground emission grows.  

%%%%%%%%%%%%%%%%%%%%%%%%%%%%%%%%%%%%%
%%%%%%%%%%%%%%%%%%%%%%%%%%%%%%%%%%%%%

It is important to remember that there is still low-redshift contamination contributing to both the mean and fluctuations, especially at longer wavelengths.  Therefore, these results should be interpreted with this in mind.  As observations improve, these results will become more reliable.

\section{CONCLUSIONS}
\label{sec:conc}

We have shown the observable signatures of high redshift populations with different values of the escape fraction $f_{esc}$ and dust content in the CIB.  It is possible to distinguish these populations through observations of the fractional anisotropy of the CIB.  The global escape fraction of high redshift galaxies is a main variable that can be probed in this way, since the angular power spectrum is dependent on it, while the mean is not.  In addition, dust will transform the SED of the galaxy, thus leaving an imprint on the fractional anisotropy.  Therefore, low values of the fractional anisotropy will be indicative of a population of stars with a high escape fraction and little dust.  This will be more noticeable at longer wavelengths.  
While observations are still difficult, improved observations could be able to distinguish between these populations.  
\\

We would like to thank Masami Ouchi and Kristian Finlator for helpful discussions.  In addition, we would like to thank Melanie Koehler and Laurent Verstraete for help with DustEM.  This work was supported by the Science and Technology Facilities Council [grant numbers ST/F002858/1 and ST/I000976/1];  the ANR
program ANR-09-BLAN-0224-02 , and The Southeast Physics Network (SEPNet). The authors acknowledge the Texas Advanced Computing Center (TACC) at The University of Texas at Austin for providing HPC resources that have contributed to the research results reported within this paper. URL: http://www.tacc.utexas.edu. This research was supported in part by the National Science Foundation through TeraGrid resources provided by TACC and NICS.

%\bibliography{mybib}

\begin{thebibliography}{63}
\expandafter\ifx\csname natexlab\endcsname\relax\def\natexlab#1{#1}\fi

\bibitem[{{Amblard} {et~al.}(2011){Amblard}, {Cooray}, {Serra}, {Altieri},
  {Arumugam}, {Aussel}, {Blain}, {Bock}, {Boselli}, {Buat},
  {Castro-Rodr{\'{\i}}guez}, {Cava}, {Chanial}, {Chapin}, {Clements}, {Conley},
  {Conversi}, {Dowell}, {Dwek}, {Eales}, {Elbaz}, {Farrah}, {Franceschini},
  {Gear}, {Glenn}, {Griffin}, {Halpern}, {Hatziminaoglou}, {Ibar}, {Isaak},
  {Ivison}, {Khostovan}, {Lagache}, {Levenson}, {Lu}, {Madden}, {Maffei},
  {Mainetti}, {Marchetti}, {Marsden}, {Mitchell-Wynne}, {Nguyen}, {O'Halloran},
  {Oliver}, {Omont}, {Page}, {Panuzzo}, {Papageorgiou}, {Pearson},
  {P{\'e}rez-Fournon}, {Pohlen}, {Rangwala}, {Roseboom}, {Rowan-Robinson},
  {Portal}, {Schulz}, {Scott}, {Seymour}, {Shupe}, {Smith}, {Stevens},
  {Symeonidis}, {Trichas}, {Tugwell}, {Vaccari}, {Valiante}, {Valtchanov},
  {Vieira}, {Vigroux}, {Wang}, {Ward}, {Wright}, {Xu}, \&
  {Zemcov}}]{Amblard/etal:2011}
{Amblard}, A., {et~al.} 2011, Nature, 470, 510

\bibitem[{{Berta} {et~al.}(2010){Berta}, {Magnelli}, {Lutz}, {Altieri},
  {Aussel}, {Andreani}, {Bauer}, {Bongiovanni}, {Cava}, {Cepa}, {Cimatti},
  {Daddi}, {Dominguez}, {Elbaz}, {Feuchtgruber}, {F{\"o}rster Schreiber},
  {Genzel}, {Gruppioni}, {Katterloher}, {Magdis}, {Maiolino}, {Nordon},
  {P{\'e}rez Garc{\'{\i}}a}, {Poglitsch}, {Popesso}, {Pozzi}, {Riguccini},
  {Rodighiero}, {Saintonge}, {Santini}, {Sanchez-Portal}, {Shao}, {Sturm},
  {Tacconi}, {Valtchanov}, {Wetzstein}, \& {Wieprecht}}]{Berta/etal:2010}
{Berta}, S., {et~al.} 2010, \aap, 518, L30+

\bibitem[{{Berta} {et~al.}(2011){Berta}, {Magnelli}, {Nordon}, {Lutz}, {Wuyts},
  {Altieri}, {Andreani}, {Aussel}, {Castaneda}, {Cepa}, {Cimatti}, {Daddi},
  {Elbaz}, {Foerster Schreiber}, {Genzel}, {Le Floc'h}, {Maiolino},
  {Perez-Fournon}, {Poglitsch}, {Popesso}, {Pozzi}, {Riguccini}, {Rodighiero},
  {Sanchez-Portal}, {Sturm}, {Tacconi}, \& {Valtchanov}}]{Berta/etal:2011}
---. 2011, \aap, 532, 49

\bibitem[{{B{\'e}thermin} {et~al.}(2012){B{\'e}thermin}, {Le Floc'h}, {Ilbert},
  {Conley}, {Lagache}, {Amblard}, {Arumugam}, {Aussel}, {Berta}, {Bock},
  {Boselli}, {Buat}, {Casey}, {Castro-Rodr{\'{\i}}guez}, {Cava}, {Clements},
  {Cooray}, {Dowell}, {Eales}, {Farrah}, {Franceschini}, {Glenn}, {Griffin},
  {Hatziminaoglou}, {Heinis}, {Ibar}, {Ivison}, {Kartaltepe}, {Levenson},
  {Magdis}, {Marchetti}, {Marsden}, {Nguyen}, {O'Halloran}, {Oliver}, {Omont},
  {Page}, {Panuzzo}, {Papageorgiou}, {Pearson}, {P{\'e}rez-Fournon}, {Pohlen},
  {Rigopoulou}, {Roseboom}, {Rowan-Robinson}, {Salvato}, {Schulz}, {Scott},
  {Seymour}, {Shupe}, {Smith}, {Symeonidis}, {Trichas}, {Tugwell}, {Vaccari},
  {Valtchanov}, {Vieira}, {Viero}, {Wang}, {Xu}, \&
  {Zemcov}}]{Bethermin/etal:2012}
{B{\'e}thermin}, M., {et~al.} 2012, \aap, 542, A58

\bibitem[{{Bouwens} {et~al.}(2010){Bouwens}, {Illingworth}, {Oesch}, {Trenti},
  {Stiavelli}, {Carollo}, {Franx}, {van Dokkum}, {Labb{\'e}}, \&
  {Magee}}]{Bouwens/etal:2010}
{Bouwens}, R.~J., {et~al.} 2010, ApJL, 708, L69

\bibitem[{{Cambr{\'e}sy} {et~al.}(2001){Cambr{\'e}sy}, {Reach}, {Beichman}, \&
  {Jarrett}}]{cambresy/etal:2001}
{Cambr{\'e}sy}, L., {Reach}, W.~T., {Beichman}, C.~A., \& {Jarrett}, T.~H.
  2001, ApJ, 555, 563

\bibitem[{{Compi{\`e}gne} {et~al.}(2010){Compi{\`e}gne}, {Flagey},
  {Noriega-Crespo}, {Martin}, {Bernard}, {Paladini}, \&
  {Molinari}}]{compiegne/etal:2010}
{Compi{\`e}gne}, M., {Flagey}, N., {Noriega-Crespo}, A., {Martin}, P.~G.,
  {Bernard}, J.-P., {Paladini}, R., \& {Molinari}, S. 2010, ApJL, 724, L44

\bibitem[{{Compi{\`e}gne} {et~al.}(2011){Compi{\`e}gne}, {Verstraete}, {Jones},
  {Bernard}, {Boulanger}, {Flagey}, {Le Bourlot}, {Paradis}, \&
  {Ysard}}]{compiegne/etal:2011}
{Compi{\`e}gne}, M., {et~al.} 2011, A\&A, 525, A103

\bibitem[{{Cooray} {et~al.}(2004){Cooray}, {Bock}, {Keatin}, {Lange}, \&
  {Matsumoto}}]{cooray/etal:2004}
{Cooray}, A., {Bock}, J.~J., {Keatin}, B., {Lange}, A.~E., \& {Matsumoto}, T.
  2004, ApJ, 606, 611

\bibitem[{{Cooray} \& {Yoshida}(2004)}]{cooray/yoshida:2004}
{Cooray}, A., \& {Yoshida}, N. 2004, MNRAS, 351, L71

\bibitem[{{Cooray} {et~al.}(2007){Cooray}, {Sullivan}, {Chary}, {Bock},
  {Dickinson}, {Ferguson}, {Keating}, {Lange}, \& {Wright}}]{cooray/etal:2007}
{Cooray}, A., {et~al.} 2007, ApJL, 659, L91

\bibitem[{{Dole} {et~al.}(2006){Dole}, {Lagache}, {Puget}, {Caputi},
  {Fern{\'a}ndez-Conde}, {Le Floc'h}, {Papovich}, {P{\'e}rez-Gonz{\'a}lez},
  {Rieke}, \& {Blaylock}}]{Dole/etal:2006}
{Dole}, H., {et~al.} 2006, \aap, 451, 417

\bibitem[{{Dwek} \& {Arendt}(1998)}]{dwek/arendt:1998}
{Dwek}, E., \& {Arendt}, R.~G. 1998, ApJL, 508, L9

\bibitem[{{Fernandez} {et~al.}(2012){Fernandez}, {Iliev}, {Komatsu}, \&
  {Shapiro}}]{fernandez/etal:2011}
{Fernandez}, E.~R., {Iliev}, I.~T., {Komatsu}, E., \& {Shapiro}, P.~R. 2012,
  \apj, 750, 20

\bibitem[{{Fernandez} \& {Komatsu}(2006)}]{fernandez/komatsu:2006}
{Fernandez}, E.~R., \& {Komatsu}, E. 2006, ApJ, 646, 703

\bibitem[{{Fernandez} {et~al.}(2010){Fernandez}, {Komatsu}, {Iliev}, \&
  {Shapiro}}]{fernandez/etal:2010}
{Fernandez}, E.~R., {Komatsu}, E., {Iliev}, I.~T., \& {Shapiro}, P.~R. 2010,
  ApJ, 710, 1089

\bibitem[{{Fernandez} \& {Shull}(2011)}]{fernandez/shull:2011}
{Fernandez}, E.~R., \& {Shull}, J.~M. 2011, ApJ, 731, 20

\bibitem[{{Fixsen} {et~al.}(1998){Fixsen}, {Dwek}, {Mather}, {Bennett}, \&
  {Shafer}}]{fixsen/etal:1998}
{Fixsen}, D.~J., {Dwek}, E., {Mather}, J.~C., {Bennett}, C.~L., \& {Shafer},
  R.~A. 1998, ApJ, 508, 123

\bibitem[{{Gorjian} {et~al.}(2000){Gorjian}, {Wright}, \&
  {Chary}}]{gorjian/wright/chary:2000}
{Gorjian}, V., {Wright}, E.~L., \& {Chary}, R.~R. 2000, ApJ, 536, 550

\bibitem[{{Greve} {et~al.}(2010){Greve}, {Wei{$\beta$}}, {Walter}, {Smail},
  {Zheng}, {Knudsen}, {Coppin}, {Kov{\'a}cs}, {Bell}, {de Breuck},
  {Dannerbauer}, {Dickinson}, {Gawiser}, {Lutz}, {Rix}, {Schinnerer},
  {Alexander}, {Bertoldi}, {Brandt}, {Chapman}, {Ivison}, {Koekemoer},
  {Kreysa}, {Kurczynski}, {Menten}, {Siringo}, {Swinbank}, \& {van der
  Werf}}]{Greve/etal:2010}
{Greve}, T.~R., {et~al.} 2010, ApJ, 719, 483

\bibitem[{{Grossan} \& {Smoot}(2007)}]{Grossan/smoot:2007}
{Grossan}, B., \& {Smoot}, G.~F. 2007, \aap, 474, 731

\bibitem[{{Hauser} {et~al.}(1998){Hauser}, {Arendt}, {Kelsall}, {Dwek},
  {Odegard}, {Weiland}, {Freudenreich}, {Reach}, {Silverberg}, {Moseley},
  {Pei}, {Lubin}, {Mather}, {Shafer}, {Smoot}, {Weiss}, {Wilkinson}, \&
  {Wright}}]{Hauser/etal:1998}
{Hauser}, M.~G., {et~al.} 1998, ApJ, 508, 25

\bibitem[{{Iliev} {et~al.}(2012){Iliev}, {Mellema}, {Shapiro}, {Pen}, {Mao},
  {Koda}, \& {Ahn}}]{iliev/etal:2011}
{Iliev}, I.~T., {Mellema}, G., {Shapiro}, P.~R., {Pen}, U.-L., {Mao}, Y.,
  {Koda}, J., \& {Ahn}, K. 2012, \mnras, 423, 2222

\bibitem[{{Jauzac} {et~al.}(2011){Jauzac}, {Dole}, {Le Floc'h}, {Aussel},
  {Caputi}, {Ilbert}, {Salvato}, {Bavouzet}, {Beelen}, {B{\'e}thermin},
  {Kneib}, {Lagache}, \& {Puget}}]{Jauzac/etal:2011}
{Jauzac}, M., {et~al.} 2011, \aap, 525, A52+

\bibitem[{{Kashlinsky}(2005)}]{kashlinsky:2005}
{Kashlinsky}, A. 2005, Phys. Rep., 409, 361

\bibitem[{{Kashlinsky} {et~al.}(2004){Kashlinsky}, {Arendt}, {Gardner},
  {Mather}, \& {Moseley}}]{kashlinsky/etal:2004}
{Kashlinsky}, A., {Arendt}, R., {Gardner}, J.~P., {Mather}, J.~C., \&
  {Moseley}, S.~H. 2004, ApJ, 608, 1

\bibitem[{{Kashlinsky} {et~al.}(2012){Kashlinsky}, {Arendt}, {Ashby}, {Fazio},
  {Mather}, \& {Moseley}}]{kashlinsky/etal:2012}
{Kashlinsky}, A., {Arendt}, R.~G., {Ashby}, M.~L.~N., {Fazio}, G.~G., {Mather},
  J., \& {Moseley}, S.~H. 2012, ApJ, 753, 63

\bibitem[{{Kashlinsky} {et~al.}(2005){Kashlinsky}, {Arendt}, {Mather}, \&
  {Moseley}}]{kashlinsky/etal:2005}
{Kashlinsky}, A., {Arendt}, R.~G., {Mather}, J., \& {Moseley}, S.~H. 2005,
  Nature, 438, 45

\bibitem[{{Kashlinsky} {et~al.}(2007{\natexlab{a}}){Kashlinsky}, {Arendt},
  {Mather}, \& {Moseley}}]{kash/etal:2007}
---. 2007{\natexlab{a}}, ApJL, 666, L1

\bibitem[{{Kashlinsky} {et~al.}(2007{\natexlab{b}}){Kashlinsky}, {Arendt},
  {Mather}, \& {Moseley}}]{kashlinskyb/etal:2007c}
---. 2007{\natexlab{b}}, ApJL, 654, L5

\bibitem[{{Kashlinsky} {et~al.}(2007{\natexlab{c}}){Kashlinsky}, {Arendt},
  {Mather}, \& {Moseley}}]{kashlinsky/etal:2007}
---. 2007{\natexlab{c}}, ApJL, 654, L1

\bibitem[{{Kashlinsky} \& {Odenwald}(2000)}]{kashlinsky/odenwald:2000}
{Kashlinsky}, A., \& {Odenwald}, S. 2000, ApJ, 528, 74

\bibitem[{{Kashlinsky} {et~al.}(2002){Kashlinsky}, {Odenwald}, {Mather},
  {Skrutskie}, \& {Cutri}}]{kashlinsky/etal:2002}
{Kashlinsky}, A., {Odenwald}, S., {Mather}, J., {Skrutskie}, M.~F., \& {Cutri},
  R.~M. 2002, ApJL, 579, L53

\bibitem[{{Komatsu} {et~al.}(2009){Komatsu}, {Dunkley}, {Nolta}, {Bennett},
  {Gold}, {Hinshaw}, {Jarosik}, {Larson}, {Limon}, {Page}, {Spergel},
  {Halpern}, {Hill}, {Kogut}, {Meyer}, {Tucker}, {Weiland}, {Wollack}, \&
  {Wright}}]{komatsu/etal:2008}
{Komatsu}, E., {et~al.} 2009, ApJS, 180, 330

\bibitem[{{Komatsu} {et~al.}(2011){Komatsu}, {Smith}, {Dunkley}, {Bennett},
  {Gold}, {Hinshaw}, {Jarosik}, {Larson}, {Nolta}, {Page}, {Spergel},
  {Halpern}, {Hill}, {Kogut}, {Limon}, {Meyer}, {Odegard}, {Tucker}, {Weiland},
  {Wollack}, \& {Wright}}]{komatsu/etal:2011}
---. 2011, ApJS, 192, 18

\bibitem[{{Lagache} {et~al.}(2007){Lagache}, {Bavouzet}, {Fernandez-Conde},
  {Ponthieu}, {Rodet}, {Dole}, {Miville-Desch{\^e}nes}, \&
  {Puget}}]{Lagache/etal:2007}
{Lagache}, G., {Bavouzet}, N., {Fernandez-Conde}, N., {Ponthieu}, N., {Rodet},
  T., {Dole}, H., {Miville-Desch{\^e}nes}, M.-A., \& {Puget}, J.-L. 2007, ApJL,
  665, L89

\bibitem[{{Lagache} \& {Puget}(2000)}]{Lagache/Puget:2000}
{Lagache}, G., \& {Puget}, J.~L. 2000, \aap, 355, 17

\bibitem[{{Lagache} {et~al.}(2000){Lagache}, {Puget}, {Abergel}, {Desert},
  {Dole}, {Bouchet}, {Boulanger}, {Ciliegi}, {Clements}, {Cesarsky}, {Elbaz},
  {Franceschini}, {Gispert}, {Guiderdoni}, {Haffner}, {Harwit}, {Laureijs},
  {Lemke}, {Moorwood}, {Oliver}, {Reach}, {Reynolds}, {Rowan-Robinson},
  {Stickel}, \& {Tufte}}]{Lagache/etal:2000}
{Lagache}, G., {et~al.} 2000, in Lecture Notes in Physics, Berlin Springer
  Verlag, Vol. 548, ISO Survey of a Dusty Universe, ed. {D.~Lemke, M.~Stickel,
  \& K.~Wilke}, 81--+

\bibitem[{{Larson}(1998)}]{larson:1998}
{Larson}, R.~B. 1998, MNRAS, 301, 569

\bibitem[{{Madau} \& {Silk}(2005)}]{madau/silk:2005}
{Madau}, P., \& {Silk}, J. 2005, MNRAS, 359, L37

\bibitem[{{Magliocchetti} {et~al.}(2003){Magliocchetti}, {Salvaterra}, \&
  {Ferrara}}]{magliocchetti/salvaterra/ferrara:2003}
{Magliocchetti}, M., {Salvaterra}, R., \& {Ferrara}, A. 2003, MNRAS, 342, L25

\bibitem[{{Marsden} {et~al.}(2009){Marsden}, {Ade}, {Bock}, {Chapin}, {Devlin},
  {Dicker}, {Griffin}, {Gundersen}, {Halpern}, {Hargrave}, {Hughes}, {Klein},
  {Mauskopf}, {Magnelli}, {Moncelsi}, {Netterfield}, {Ngo}, {Olmi}, {Pascale},
  {Patanchon}, {Rex}, {Scott}, {Semisch}, {Thomas}, {Truch}, {Tucker},
  {Tucker}, {Viero}, \& {Wiebe}}]{Marsden/etal:2009}
{Marsden}, G., {et~al.} 2009, ApJ, 707, 1729

\bibitem[{{Matsumoto} {et~al.}(2005){Matsumoto}, {Matsuura}, {Murakami},
  {Tanaka}, {Freund}, {Lim}, {Cohen}, {Kawada}, \&
  {Noda}}]{matsumoto/etal:2005}
{Matsumoto}, T., {et~al.} 2005, ApJ, 626, 31

\bibitem[{{Matsumoto} {et~al.}(2011){Matsumoto}, {Seo}, {Jeong}, {Lee},
  {Matsuura}, {Matsuhara}, {Oyabu}, {Pyo}, \& {Wada}}]{matsumoto/etal:2011}
---. 2011, ApJ, 742, 124

\bibitem[{{Matsuura} {et~al.}(2011){Matsuura}, {Shirahata}, {Kawada},
  {Takeuchi}, {Burgarella}, {Clements}, {Jeong}, {Hanami}, {Khan}, {Matsuhara},
  {Nakagawa}, {Oyabu}, {Pearson}, {Pollo}, {Serjeant}, {Takagi}, \&
  {White}}]{matsuura/etal:2011}
{Matsuura}, S., {et~al.} 2011, ApJ, 737, 2

\bibitem[{{Miville-Desch{\^e}nes} {et~al.}(2002){Miville-Desch{\^e}nes},
  {Lagache}, \& {Puget}}]{Miville-Deschenes+2002}
{Miville-Desch{\^e}nes}, M.-A., {Lagache}, G., \& {Puget}, J.-L. 2002, \aap,
  393, 749

\bibitem[{{Odegard} {et~al.}(2007){Odegard}, {Arendt}, {Dwek}, {Haffner},
  {Hauser}, \& {Reynolds}}]{odegard/etal:2007}
{Odegard}, N., {Arendt}, R.~G., {Dwek}, E., {Haffner}, L.~M., {Hauser}, M.~G.,
  \& {Reynolds}, R.~J. 2007, ApJ, 667, 11

\bibitem[{{Odenwald} {et~al.}(2003){Odenwald}, {Kashlinsky}, {Mather},
  {Skrutskie}, \& {Cutri}}]{odenwald/etal:2003}
{Odenwald}, S., {Kashlinsky}, A., {Mather}, J.~C., {Skrutskie}, M.~F., \&
  {Cutri}, R.~M. 2003, ApJ, 583, 535

\bibitem[{{P{\'e}nin} {et~al.}(2011){P{\'e}nin}, {Lagache}, {Noriega-Crepo},
  {Grain}, {Miville-Desch{\^e}nes}, {Ponthieu}, {Martin}, {Blagrave}, \&
  {Lockman}}]{Penin/etal:2011}
{P{\'e}nin}, A., {et~al.} 2011, \aap, 543, 123

\bibitem[{{Planck Collaboration} {et~al.}(2011){Planck Collaboration}, {Ade},
  {Aghanim}, {Arnaud}, {Ashdown}, {Aumont}, {Baccigalupi}, {Balbi}, {Banday},
  {Barreiro}, \& et~al.}]{Planck:2011}
{Planck Collaboration} {et~al.} 2011, \aap, 536, A18

\bibitem[{{Pyo} {et~al.}(2012){Pyo}, {Matsumoto}, {Jeong}, \&
  {Matsuura}}]{pyo/etal:2012}
{Pyo}, J., {Matsumoto}, T., {Jeong}, W.-S., \& {Matsuura}, S. 2012, ArXiv
  1202.4049

\bibitem[{{Riechers} {et~al.}(2010){Riechers}, {Capak}, {Carilli}, {Cox},
  {Neri}, {Scoville}, {Schinnerer}, {Bertoldi}, \& {Yan}}]{riechers/etal:2010}
{Riechers}, D.~A., {et~al.} 2010, ApJL, 720, L131

\bibitem[{{Robertson} {et~al.}(2010){Robertson}, {Ellis}, {Dunlop}, {McLure},
  \& {Stark}}]{Robertson/etal:2010}
{Robertson}, B.~E., {Ellis}, R.~S., {Dunlop}, J.~S., {McLure}, R.~J., \&
  {Stark}, D.~P. 2010, Nature, 468, 49

\bibitem[{{Salpeter}(1955)}]{salpeter:1955}
{Salpeter}, E.~E. 1955, ApJ, 121, 161

\bibitem[{{Salvaterra} \& {Ferrara}(2003)}]{salvaterra/ferrara:2003}
{Salvaterra}, R., \& {Ferrara}, A. 2003, MNRAS, 339, 973

\bibitem[{{Santos} {et~al.}(2002){Santos}, {Bromm}, \&
  {Kamionkowski}}]{santos/bromm/kamionkowski:2002}
{Santos}, M.~R., {Bromm}, V., \& {Kamionkowski}, M. 2002, MNRAS, 336, 1082

\bibitem[{{Thompson} {et~al.}(2007{\natexlab{a}}){Thompson}, {Eisenstein},
  {Fan}, {Rieke}, \& {Kennicutt}}]{thompson/etal:2007a}
{Thompson}, R.~I., {Eisenstein}, D., {Fan}, X., {Rieke}, M., \& {Kennicutt},
  R.~C. 2007{\natexlab{a}}, ApJ, 657, 669

\bibitem[{{Thompson} {et~al.}(2007{\natexlab{b}}){Thompson}, {Eisenstein},
  {Fan}, {Rieke}, \& {Kennicutt}}]{thompson/etal:2007b}
---. 2007{\natexlab{b}}, ApJ, 666, 658

\bibitem[{{Totani} {et~al.}(2001){Totani}, {Yoshii}, {Iwamuro}, {Maihara}, \&
  {Motohara}}]{totani/etal:2001}
{Totani}, T., {Yoshii}, Y., {Iwamuro}, F., {Maihara}, T., \& {Motohara}, K.
  2001, ApJL, 550, L137

\bibitem[{{Viero} {et~al.}(2009){Viero}, {Ade}, {Bock}, {Chapin}, {Devlin},
  {Griffin}, {Gundersen}, {Halpern}, {Hargrave}, {Hughes}, {Klein},
  {MacTavish}, {Marsden}, {Martin}, {Mauskopf}, {Moncelsi}, {Negrello},
  {Netterfield}, {Olmi}, {Pascale}, {Patanchon}, {Rex}, {Scott}, {Semisch},
  {Thomas}, {Truch}, {Tucker}, {Tucker}, \& {Wiebe}}]{Viero/etal:2009}
{Viero}, M.~P., {et~al.} 2009, ApJ, 707, 1766

\bibitem[{{Wright}(2001)}]{wright:2001}
{Wright}, E.~L. 2001, ApJ, 553, 538

\bibitem[{{Wright}(2004)}]{wright:2004}
---. 2004, New AR, 48, 465

\bibitem[{{Wright} \& {Reese}(2000)}]{wright/reese:2000}
{Wright}, E.~L., \& {Reese}, E.~D. 2000, ApJ, 545, 43

\end{thebibliography}

\end{document}